\documentclass{PoS}

\usepackage{amsmath, amsthm, amssymb}
\usepackage{natbib}
\usepackage{amssymb}
\usepackage{graphicx}
\usepackage{enumitem}

\newcommand{\g}{\gamma}
\newcommand{\tbh}{\hat{t}_b}

\title{New insights on the duration distribution of long GRBs from Collapsars}

\ShortTitle{Duration distribution of Collapsars}

\author{\speaker{O. Bromberg}$^a$, E. Nakar$^{ab}$, T. Piran$^a$ \& R. Sari$^a$\\
        \llap{$^a$}Racah Institute of Physics, The Hebrew University, 91904 Jerusalem, Israel\\
        \llap{$^{ab}$}The Raymond and Berverly Sackler School of Physics and Astronomy,
 Tel Aviv University, 69978 Tel Aviv, Israel\\
        E-mail: \email{omerb@phys.huji.ac.il}}

\abstract{
According the Collapsar model long gamma-ray bursts (LGRBs)
involve relativistic jets that puncture the envelope
of a collapsing star, and produced the $\gamma$-rays after
they break out.
This model provides a theoretical framework for the well known
association between LGRBs and massive stars.
However although this association is supported by a wealth of
observations, to this date there is no direct observational evidence for the
emergence of the jet from the star. In other words there is no
direct evidence for the Collapsar model.
Here we show that a distinct signature of
the Collapsar model is the appearance  of a plateau in the duration
distribution of the prompt GRB emission at times much shorter than
the typical breakout time of the jet.
This plateau is evident in the data of all major GRB satellites,
and provides a direct evidence supporting the Collapsar model.
It also enables us to place limits on the sizes and masses of LGRB progenitors;
suggests the existence of a large population of choked (failed)
GRBs; and indicates that the 2 s duration commonly used to separate Collapsars
and non-Collapsars holds for BATSE and possibly Fermi GBM GRBs, but it
is inconsistent with the duration distributions
of {\it Swift} GRBs.}

\FullConference{The Extreme and Variable High Energy Sky - extremesky2011,\\
		September 19-23, 2011\\
		Chia Laguna (Cagliari), Italy}

\begin{document}
\section{Introduction}

The connection of long GRBs (LGRBs) to collapsing massive stars
is strongly supported by a wealth of observational evidence.
The most notable evidence is the association of half
a dozens GRBs with spectroscopically confirmed broad-line Ic
supernovae (SNe), and the identification of "red bumps" in the
afterglows of about two dozens more, which shows a photometric
evidence of underlying SNe. The model that provides the theoretical framework of
this association is known as the Collapsar model
\citep{MacWoos99,MacFadyen01}. According to this model, following
the core collapse of a massive star, a bipolar jet is launched at
the center of the star. The jet drills through the stellar envelope
and breaks out of the surface before producing the observed
$\gamma$-rays. However, although this model is supported
indirectly by the LGRB-SN association, to this date we could not
identify a clear direct observational imprint of the jet-envelope
interaction, thus there is no direct confirmation yet of the
Collapsar model.

Here we show that under very general
conditions the time that the jet spends drilling through the star
leads to a plateau in the duration distribution of the GRBs at times much
shorter than the breakout time of the jet. This plateau
exists in the duration
distribution of all major GRB satellites, and provides a strong
observational support for the Collapsar model. An
analysis of this plateau also
(i) supports the hypothesis of compact
stellar progenitors, (ii) implies the existence of  a large population
of chocked jets that fail to break out of their progenitor stars and
(iii) enables us to determine statistically the fraction of
non-Collapsars from the total GRB sample and the threshold duration that
separates Collapsars from non-Collapsars.
Specifically it
shows that this time 
differs from one satellite to the other.

\section{The propagation of a jet in the stellar envelope}

The jet propagates in the star by pushing the stellar material in front
of it, leading to the formation of a ``head" of shocked matter at its front.
The head propagates at sub relativistic velocities along most
of the star \citep{Matzner03,Zhang03,Morsony07,Mizuta09,B11a},
even though the jet is ejected at relativistic velocities.
Therefore as long as the head is inside the star it dissipates most of the jet
energy, and it needs to be constantly supported by the relativistic jet material
to propagate.
This implies that there is a minimal amount of time, $t_b$, that the jet engine
needs to operate to get a successful jet breakout \citep{B11b}:
\begin{equation}\label{eq:tB_GRB}
t_b \simeq15~{\rm s} \cdot \left( \frac{ L_{iso}} {10^{51} {\rm~
erg/s}}\right)^{-1/3} \left (\frac{\theta}{10^\circ}\right)^{2/3}
\left (  \frac{R_{*}} {5 R_\odot}\right)^{2/3} \left
(\frac{M_{*}}{15M_\odot}\right)^{1/3},
\end{equation}
where $ L_{iso}$ is the isotropic equivalent jet luminosity,
$\theta$ is the jet half opening angle and we have used typical
values for a long GRB. $ R_{*}$ and $M_{*}$ are the radius and the
mass of the progenitor star, where we normalize their value
according to the typical radius and mass inferred from observations
of the few supernovae (SNe) associated with long GRBs.
If the activity time of the engine
$t_e < t_b$ the jet fails to escape and a regular long GRB is not observed.

After the jet breaks out of the star it expands and
produces the observed $\gamma$-ray
emission  at large distances from the stellar surface.
In most GRB models \citep[e.g.][]{SariPiran97} the observed
duration of the GRB, $t_\gamma$, reflects the activity time of the central
engine after the break out of the jet
\begin{equation}
\label{tgamma} t_\gamma = t_{e} -  t_b .
\end{equation}
The distribution of $t_\gamma$ is therefore a convolution of  the
distributions of the engine activity time and the breakout time
combined with cosmological redshift effects.

\section{The duration distribution of Collapsars}\label{sec theory}

Under very general conditions, Eq. \ref{tgamma} results in a flat
distribution of $t_\gamma$ at durations significantly shorter than
the typical breakout time. A simple way to show that is by considering
a single value
of $t_b$ for all Collapsars and ignoring, for simplicity, cosmological redshifts and
detector threshold effects. The probability that a GRB has a
duration $t_\gamma$ equals, in this case,  to the probability that
the engine operating time is $t_\gamma+t_b$, i.e.
\begin{equation}\label{eq:p_g}
p_\gamma(t_\gamma) dt_\gamma=p_{e}(t_b+t_\gamma) dt_\g,
\end{equation}
where $p_\gamma$ is the probability
distribution of the observed durations and $p_{e}$ is the probability
distribution of the engine operating times.
It is clear that $p_e(t_b+t_\g) \approx p_e(t_b) =$ const for any $t_\g\ll t_b$.
Moreover, if $p_e(t_e)$ is a smooth
function that does not vary rapidly in the vicinity of $t_e \approx
t_b$, over a duration of the scale of $t_b$, then the constant
distribution is extended up to times $t_\g\lesssim t_b$. In the case
of interest  $t_b$ and $t_e$ are determined by different regions of
the star: the breakout time is set by the density and radius of the
stellar envelope at radii $>10^{10}$ cm, while  $t_e$ is determined
by the stellar core properties at radii $<10^8$ cm. The core and the
envelope are weakly coupled \citep{Crowther07} and the engine is
unaware whether the jet has broken out or not. Therefore, it is
reasonable to expect that $p_e(t_e)$ is smooth in the vicinity of $t_e
\approx t_b$ and $p_\gamma (t_\gamma) \approx$ const for $t_\gamma
\lesssim t_b$. In the opposite limit, where $t_b \ll t_\g$, then
$t_\g \approx t_e$. Together eq. \ref{eq:p_g} reads:
\begin{equation}\label{tg}
p_\gamma (t_\gamma)  \approx \left\{
\begin{array}{cc}
  p_e(t_b) &  t_\gamma \lesssim t_b   \\
  p_e(t_\gamma) &   t_\gamma \gg t_b
\end{array}
\right. ,
\end{equation}

Remarkably, as we show in \citet{B11c}, the plateau exists also in the more general case, when
$t_b$ varies and the redshift distribution and detector thresholds are considered. The
constant value of $t_b$, in this case, should be replaced by a ``typical" breakout
time, $\tbh$, modulo redshift corrections.
This plateau  in the GRB duration distribution is
independent of specific details of the relevant distributions and in
particular of the details of $p_e$. It is a direct prediction of the
Collapsar model that  follows immediately from Eq. \ref{tgamma}.
Given an average redshift $\overline {(1+z)} \sim 3$ and  Eq.
\ref{eq:tB_GRB}, we expect the effective (redshift corrected)
typical breakout time to be of order 50 s.
At long observed durations, $t_\g\gg\tbh$,  the distribution $p_\g$
is determined by a convolution of $p_e$ and the distributions of the breakout times and
bursts' redshifts. However $p_\g$ cannot be flatter than $p_e$, otherwise
$p_e$ would dominate the distribution.
Therefore, an
extrapolation of $p_\g(t_\g \gg \tbh)$ to durations shorter than
$\tbh$ provides a lower limit to the number of events with $t_e<\tbh$.
Namely, it is an
estimate of the minimal number of choked bursts.
At very short durations (less than 2 s) an additional
population of shorter and harder GRBs (SGRBs) appears \citep{Kouveliotou93}. It is well
established that most SGRBs are not associated with death of massive
stars
\citep[and references there in]{Nakar07},
namely they are non-Collapsars, and the
above argument of a flat distribution doesn't apply to them.  Therefore, when considering
the overall  burst duration distribution we expect a flat section
for durations significantly shorter than 50 s down to the duration
where these non-Collapsars dominate.

\begin{figure}
\vspace{-0.8cm}
\includegraphics[width=4in]{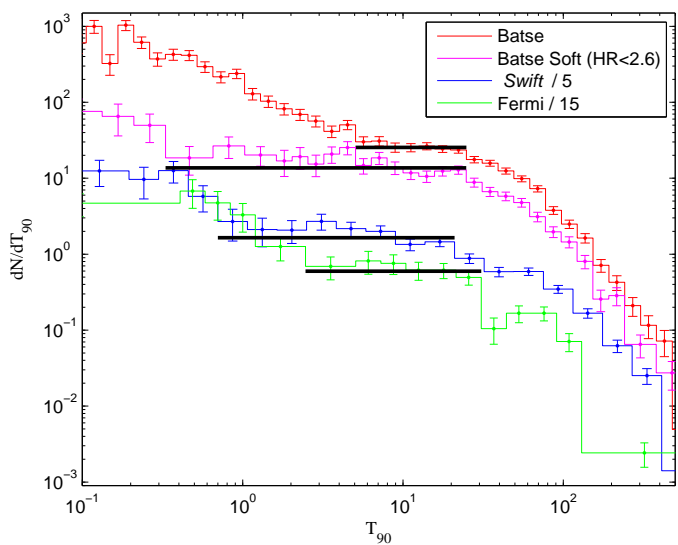}
  \caption{The $T_{90}$ distribution, $dN/dT_{90}$, of BATSE (red), {\it Swift} (blue) and Fermi GBM (green) GRBs.
    Also plotted is the distribution of the soft ($HR< 2.6$) BATSE bursts (magenta).
    For clarity the {\it Swift} values are divided by a factor of 5 and the Fermi GBM by 15.
    At shorter times the sample is dominated by non-Collapsars.
    Note that the quantity $dN/dT$  is depicted and
    not ${dN}/{d\log T}$ as traditionally shown in such plots \citep[e.g.,][]{Kouveliotou93}.
    The black lines show the best fitted flat interval in each data set:
    $5-25$ s (BATSE), $0.7-21$ s ({\it Swift}),  and $2.5-31$ s (Fermi). The upper limits of this range
    indicate a typical breakout time of a few
    dozens seconds, in agreement with the prediction of the Collapsar model.
    The distribution at times $\gtrsim 100$ s can be fitted by a power law with an  index
    $-4<\alpha<-3$.
    The soft BATSE bursts  show a considerably longer plateau
    ($0.4-25$ s), indicating that most of the soft short bursts are in fact
    Collapsars \citep{B11c}.}
\label{fig.T90_T_tot}
\end{figure}


\section{The observed distribution of the prompt GRB durations}

The observed duration of a GRB is characterized  by
$T_{90}\approx t_\gamma(1+z)$,  during which  $90\%$ of the fluence is
accumulated. Fig. 1 depicts  the observed distribution of $T_{90}$
for the three major GRB satellites:
BATSE\footnote{http://swift.gsfc.nasa.gov/docs/swift/archive/grb\_table},
{\it
Swift}\footnote{http://gammaray.msfc.nasa.gov/batse/grb/catalog/current/}
and Fermi GBM \footnote{http://lyra.berkeley.edu/grbox/grbox.php}.
Note that we plot the quantity $p_\gamma(T_{90})=dN/dT$ and
not ${dN}/{d\log T}$ traditionally shown in such
plots \citep{Kouveliotou93}. As predicted, a plateau around 2-30 s
is clearly seen in all distributions. The extent of the plateau
varies slightly from one detector to another. This is expected given
the different detection threshold sensitivities in different energy
windows (see below). The regions marked in solid bold lines in Fig.
1 are consistent with a flat distribution with a $\chi^2$ per degree
of freedom of $0.6, 1.3, 0.7$ for BATSE,  {\it Swift} and Fermi
respectively \citep[see][for details]{B11c}.
At the high end of the plateau the $T_{90}$ distribution
decreases rapidly and can be fitted at long durations ($>$100 s) by
a power law with an index, $\alpha$, in the range $-4<\alpha<-3$.

To test our hypothesis that the plateau in the duration distribution
indeed reflects the distribution of Collapsars and is not, for example,
a consequence of adding two separate distributions: one decreasing
(SGRBs) and one increasing (LGRBs), we also plot the
distribution, dN/dT,  of soft BATSE bursts (magenta). Since non-Collapsars are
harder on average \citep{Kouveliotou93}, we expect this sample to be less contaminated and the
plateau to extend to shorter durations than in the whole BATSE
sample. Considering  the soft bursts as bursts with hardness
ratio\footnote{HR is the fluence ratio between BATSE channels 3
(100-300 keV) and channel 2 (50-100 keV).}, $HR<2.6$, the median
value of bursts with $T_{90}>5$ s, we find that the plateau in this
sample extends from 25 s down to $0.4$ s (1.3 $\chi^2$/dof),
an order of magnitude lower than the
original 5 s in the complete sample. This lands a strong
support to the conclusion  that the observed flat distribution is
indeed indicating on the Collapsar origin of the population. It also
implies that $HR$ is a good indicator that effectively filters out a
large number of non-Collapsars from the BATSE GRB sample.

\section{The fraction of non-Collapsar SGRBs}

At short durations, the GRB distribution is dominated by the
non-Collapsars SGRBs
\citep[and references there in]{Nakar07}.
These bursts are hard to classify since all their hard energy
properties largely overlap with those of the Collapsars \citep{Nakar07}.
The  least overlap is in the duration distributions and hence, traditionally a
burst is classified as a non-Collapsar if $T_{90} < 2$ s. Even
though this criterion is based on the duration distribution of BATSE
it is widely used for bursts detected by all satellites.

The fraction of Collapsars at short durations can be estimated by
extrapolating the plateau in the duration distribution to short times.
Since there
is also an overlap with the SGRBs at long durations, the real heights
of the plateaus are somewhat lower than what is shown in fig. 1.
In \citet{B11d} we make a joint fit of the non-Collapsars and the Collapsars
by modeling the duration distribution of the non-Collapsars as well.
The fitted distributions are then used to calculate
the fraction $f$ of the non-Collapsar from the total
number of observed GRBs as a function of $T_{90}$. This fraction
represents the probability that an observed GRB with
a given duration in not a Collapsar.
Fig. 2 depicts $f$ as function of $T_{90}$ for
BATSE, {\it Swift} and Fermi GBM samples.
Table 1 gives $T_{90}$ at several selected values of this fraction.
It can clearly be seen that using
$T_{90}<2$ s as a method to identify non-Collapsars
works reasonably well for BATSE and is marginal in the case of Fermi,
but in {\it Swift}
data it results in
a large over estimate of the number of non-Collapsars.
Adopting, for example, $f=0.5$ as the threshold fraction
that separates the two populations, we find that for BATSE
the transition occurs at $\sim3.5$ s, for {\it Swift} it occurs
at $\sim0.6$ s and  for Fermi at $\sim 1.7$ s (Note, however, that the errors
in the Fermi data are quite large due to lack of statistics).
This shift in the
transition time is expected since non-Collapsars are on average
harder than Collapsars and different detectors have different energy detection
windows. BATSE has the hardest detection window, making it
relatively more sensitive to non-Collapsar GRBs. {\it Swift} has the
softest detection window making it relatively more sensitive to
Collapsar GRBs.

 \begin{figure}
 \vspace{-0.8cm}
\includegraphics[width=4in]{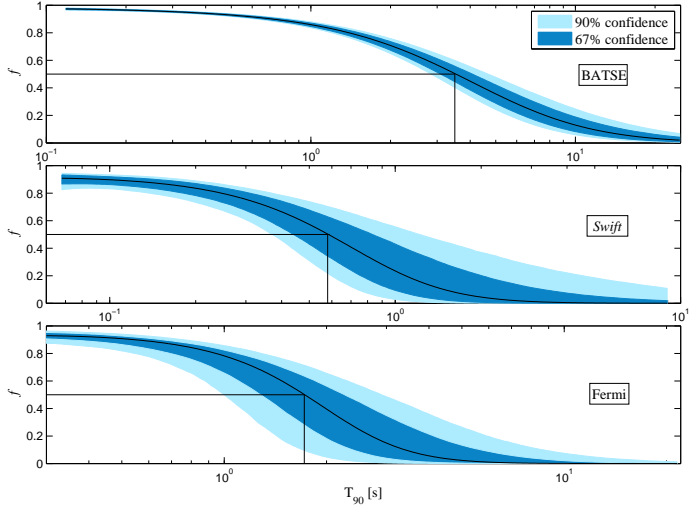}
  \caption{The fraction $f$ of non-Collapsars from the total
  number of observed GRBs as a function of the observed duration time, $T_{90}$,
  in the (from top to bottom) BATSE, {\it Swift} \& Fermi GBM samples.
  The shaded regions represent 67\% and 90\% confidence limits of $f$.
  Also plotted are the $T_{90}$ for which $f=0.5$ at each data set, the numerical
  values are given in table 1 \citep[][]{B11d}.}
\end{figure}

\begin{table}[h]
  \centering
\begin{tabular} {c c c c c c c}
 \hline\hline
  Satellite && $T_{90}(f=0.5)[s]$ &&& $T_{90}(f=0.7)[s]$ & $T_{90}(f=0.9)[s]$ \\
  \hline
BATSE && $3.5\pm0.5$ &&& $2\pm0.2$ & $0.7\pm0.06$\\
{\it Swift} && $0.6_{-0.14}^{+0.2}$ &&& $0.36\pm0.09$ & $0.1^{+0.05}_{-0.04}$$^\dag$\\
Fermi && $1.7_{-0.5}^{+0.6}$ &&& $1.2\pm0.3$ & $0.5^{+0.1}_{-0.2}$$^\dag$\\
\hline
 \multicolumn{7}{l}{\footnotesize $^\dag$ We restrict the analysis of {\it Swift} and Fermi data to
 $T_{90}>0.06$ and $0.3$ s respectively due to lack of statistics.}
 \end{tabular}
\end{table}

\vspace{-0.4cm}
\section{Discussion}

The observed plateaus in all three duration distributions, and most
notably in the distribution of soft BATSE bursts, provide a
direct support for the Collapsars model for LGRBs. An inspection of
different regions of the observed temporal distribution (Fig. 1),
under the interpretation of the plateau as an imprint of the time it
takes the jet to break out of the envelope, provides further
important information.
\begin{enumerate}[itemsep=2pt,parsep=2pt,leftmargin=0.5cm]
\item{
The end of the plateau and the decrease in the number of GRBs at
long durations, allow us to estimate the typical time it takes a jet to
breakout of the progenitor's envelope. All three distributions are
flat below $\sim 10$ s in the GRB frame, implying that $\hat t_b
\sim$ a few dozen seconds. This fits nicely with the canonical
GRB parameters used in Eq. \ref{eq:tB_GRB}, and provides another
indication that the stellar progenitors of Collapsars must be compact
\citep{Matzner03}.}

\item{
The distribution at long durations
can be used to set a lower limit on the number of chocked jets,
by extrapolating the slope at $T_{90}\gg \tbh$
to durations shorter than $\tbh$ (see section \ref{sec theory}).
At durations $T_{90}\gtrsim 100$ s $p_\gamma$ can be
fitted by a power law, $p_\gamma \propto T_{90}^{\alpha}$
with $-4<\alpha<-3$. Extrapolating to $T_{90} < \hat t_b$ we find that
if $p_e$ continues with this power law to $t_e < \tbh$ there are
significantly more chocked GRBs than long ones. For example, even if we
extrapolate  this distribution only down to $t_e = \hat t_b/2$ there are $\sim10$
times more chocked GRBs than long GRBs.
This prediction is consistent
with the suggestion that shock breakout from these choked GRBs
produces low luminosity smooth and soft GRBs
\citep[][and references there in]{Nakar11}.
Indeed the rate of such low luminosity
GRBs is far larger than that of regular long GRBs
\citep{Soderberg06}.}

\item{At short durations, the duration distribution is dominated
by non-Collapsars SGRBs. These bursts are hard to classify and are commonly
defined by their shorter duration.
Our analysis shows that putting the dividing line between Collapsars and
non-Collapsars at 2 s is statistically reasonable for BATSE,
and possibly also for Fermi bursts.
However, it is clearly wrong to do so for {\it Swift}.
We also calculate the probability as a function of $T_{90}$ that a burst with a given
$T_{90}$ is not produced by a Collapsar.}

\item{While the difference in the lower limit of the plateaus is
understood qualitatively
as it depends on the spectral range of the detector,
the variance in the upper limit is less
obvious. It may reflect various selection effects in triggering
algorithms.
A more interesting possibility is that it reflects a physical origin,
e.g. different satellites probing populations with different
$\hat t_b$.  This could be explored when the statistical sample of
Fermi GBM becomes sufficiently large.
}

\end{enumerate}




\begin{thebibliography}{99}
\expandafter\ifx\csname
natexlab\endcsname\relax\def\natexlab#1{#1}\fi



\bibitem[{{Bromberg} {et~al.}(2011{\natexlab{a}}){Bromberg}, {Nakar}, {Piran},
  \& {Sari}}]{B11a}
{Bromberg}, O., {Nakar}, E., {Piran}, T., \& {Sari}, R.
2011{\natexlab{a}},
  Apj , 740, 100

\bibitem[{{Bromberg} {et~al.}(2011{\natexlab{b}}){Bromberg}, {Nakar}, \&
  {Piran}}]{B11b}
{Bromberg}, O., {Nakar}, E., \& {Piran}, T. 2011{\natexlab{b}},
ApjL, 739, L55

\bibitem[{{Bromberg} {et~al.}(2011{\natexlab{c}}){Bromberg}, {Nakar}, \&
  {Piran}\& {Sari}}]{B11c}
{Bromberg}, O., {Nakar}, E., {Piran}, T., \& {Sari}, R. 2011{\natexlab{c}}, arXiv:1111.2990

\bibitem[{{Bromberg} {et~al.}(2011{\natexlab{d}}){Bromberg}, {Nakar}, \&
  {Piran}\& {Sari}}]{B11d}
{Bromberg}, O., {Nakar}, E., {Piran}, T., \& {Sari}, R. 2011{\natexlab{d}}, in
preperation


\bibitem[{{Crowther}(2007)}]{Crowther07}
{Crowther}, P.~A. 2007, ARA\&A, 45, 177







\bibitem[{{Kouveliotou} {et~al.}(1993){Kouveliotou}, {Meegan}, {Fishman},
  {Bhat}, {Briggs}, {Koshut}, {Paciesas}, \& {Pendleton}}]{Kouveliotou93}
{Kouveliotou}, C., {Meegan}, C.~A., {Fishman}, G.~J., {Bhat}, N.~P.,
{Briggs},
  M.~S., {Koshut}, T.~M., {Paciesas}, W.~S., \& {Pendleton}, G.~N. 1993, ApjL,
  413, L101



\bibitem[{{MacFadyen} \& {Woosley}(1999)}]{MacWoos99}
{MacFadyen}, A.~I., \& {Woosley}, S.~E. 1999, Apj, 524, 262

\bibitem[{{MacFadyen} {et~al.}(2001){MacFadyen}, {Woosley}, \&
  {Heger}}]{MacFadyen01}
{MacFadyen}, A.~I., {Woosley}, S.~E., \& {Heger}, A. 2001, Apj,
550, 410

\bibitem[{{Matzner}(2003)}]{Matzner03}
{Matzner}, C.~D. 2003, MNRAS, 345, 575


\bibitem[{{Mizuta} \& {Aloy}(2009)}]{Mizuta09}
{Mizuta}, A., \& {Aloy}, M.~A. 2009, Apj, 699, 1261

\bibitem[{{Morsony} {et~al.}(2007){Morsony}, {Lazzati}, \&
  {Begelman}}]{Morsony07}
{Morsony}, B.~J., {Lazzati}, D., \& {Begelman}, M.~C. 2007, Apj,
665, 569

\bibitem[{{Nakar}(2007)}]{Nakar07}
{Nakar}, E. 2007, physrep, 442, 166

\bibitem[{{Nakar} \& {Sari}(2011)}]{Nakar11}
{Nakar}, E., \& {Sari}, R. 2011, ArXiv e-prints, 1106.2556




\bibitem[{{Sari} \& {Piran}(1997)}]{SariPiran97}
{Sari}, R., \& {Piran}, T. 1997, Apj, 485, 270

\bibitem[{{Soderberg} {et~al.}(2006){Soderberg}, {Kulkarni}, {Nakar}, {Berger},
  {Cameron}, {Fox}, {Frail}, {Gal-Yam}, {Sari}, {Cenko}, {Kasliwal},
  {Chevalier}, {Piran}, {Price}, {Schmidt}, {Pooley}, {Moon}, {Penprase},
  {Ofek}, {Rau}, {Gehrels}, {Nousek}, {Burrows}, {Persson}, \&
  {McCarthy}}]{Soderberg06}
{Soderberg}, A.~M., {et~al.} 2006, Nature, 442, 1014




\bibitem[{{Zhang} {et~al.}(2003){Zhang}, {Woosley}, \& {MacFadyen}}]{Zhang03}
{Zhang}, W., {Woosley}, S.~E., \& {MacFadyen}, A.~I. 2003, Apj,
586, 356

\end{thebibliography}
\end{document}